\documentstyle[12pt]{article}
\def\ref{\parskip=0pt\par\noindent\hangindent1truecm}
\begin{document}
\null
\vspace{1cm}
\centerline{\huge\bf The Double Quasar Q2138-431:}
\vspace{0.7cm}
\centerline{\huge\bf Lensing by a Dark Galaxy?}
\vspace{1.7cm}

\centerline{\large\bf M.R.S. Hawkins$^{1}$, D. Clements$^{2}$, J.W. Fried$^{3}$,
A.F. Heavens$^{4}$,}
\vspace{0.7cm}
\centerline{\large\bf P. V\'{e}ron$^{5}$, E.M. Minty$^4$, P. van der Werf$^{6}$
} 
\vspace{1.7cm}

\noindent
{\it $^{1}$ Royal Observatory, Blackford Hill, Edinburgh
EH9 3HJ, Scotland}\\
{\it $^{2}$ European Southern Observatory, Karl-Schwarzschild-Str. 2,
85748 Garching bei M\"{u}nchen, Germany}\\
{\it $^{3}$ Max-Planck-Institut f\"{u}r Astronomie, K\"{o}nigstuhl 17,
D-69117 Heidelberg, Germany}\\
{\it $^{4}$ Department of Astronomy, University of Edinburgh,
Blackford Hill, Edinburgh EH9 3HJ, Scotland}\\
{\it $^{5}$ Observatoire de Haute-Provence(CNRS),
F-04870 Saint-Michel l'Observatoire, France}\\
{\it $^{6}$ Sterrewacht Leiden, Postbus 9513, 2300 RA Leiden,
The Netherlands}\\
\vspace{0.7cm}

\vspace{2.5cm}

\centerline{\large\bf ABSTRACT}
\vspace{1cm}

  We report the discovery of a new gravitational lens candidate Q2138-431AB,
comprising two quasar images at a redshift of 1.641 separated by 4.5 arcsecs.
The spectra of the two images are very similar, and the redshifts agree
to better than 115 km.sec$^{-1}$.  The two images have magnitudes
$B_J = 19.8$ and $B_J = 21.0$, and in spite of a deep search and image
subtraction procedure, no lensing galaxy has been found with $R < 23.8$.
Modelling of the system configuration implies that the mass-to-light ratio
of any lensing galaxy is likely to be around $1000 M_{\odot}/L_{\odot}$,
with an absolute lower limit of $200 M_{\odot}/L_{\odot}$ for an 
Einstein-de Sitter universe.  We conclude that
the most likely explanation of the observations is gravitational lensing by
a dark galaxy, although it is possible we are seeing a binary quasar.

\newpage

\noindent
{\Large\bf 1. INTRODUCTION}
\vspace{0.5cm}

  The first secure example of a gravitational lens (Q0957+561) was discovered by
Walsh {\it et al.} (1979), and comprised two images of a quasar at redshift 1.41
lensed by a bright cluster galaxy at $z = 0.36$.  Since then many
manifestations of gravitational lensing have been observed, including
multiply lensed quasars, giant arcs around galaxy clusters, and distortions
of the distant galaxy distribution.  Historically, systems comprising a pair of
quasar images have always had a special significance in the catalogue of lensing
phenomena because of the simplicity of the geometry, and the plausibility of
using them to measure the Hubble constant (Refsdal, 1964).  There are at present
8 wide separation ($> 2''$) two component lens candidates known, but progress
towards finding a value of the Hubble constant has been slow for several
reasons.  These include lack of high quality light curves over a sufficiently
long period of time, uncertainty of the lensing geometry, the effects of
microlensing and failure to find the lensing galaxy.  In this paper we report
the discovery of a new gravitational lens candidate which highlights some of
these problems.\\

  This ninth wide separation system was discovered as part of a systematic
survey for lens candidates.  It has a separation of 4.5$''$, and the two
components had $B$ magnitudes of 19.8 and 21.0 in 1995.  There is extensive
archival photometry of the system over 20 years, and it appears to be clear
of any nearby galaxy concentrations.  The two images are strongly variable,
but as for all but two of the other known systems the lensing galaxy has not
so far been detected.  The large mass-to-light ratios of the order of several
hundred to a thousand implied by these non-detections have prompted several
authors to speculate about the possibility of `dark galaxies'.  In this paper
we report a variety of observations of the new system, and conclude that in
this case too the most probable explanation is that the quasar is being
gravitationally lensed by a dark galaxy.\\

\vspace{2cm}

\noindent
{\Large\bf 2. OBSERVATIONS}

\vspace{0.5cm}

\noindent
{\bf 2.1 The Lens Survey}

\vspace{0.3cm}

  The survey for gravitational lenses was carried out in the ESO/SERC field 287
centred on 21h 28m, --45$^{\circ}$ (1950).  Extensive plate material from the
UK Schmidt telescope exists in this field, which has formed the basis for the 
large scale quasar survey and monitoring programme of Hawkins \& V\'{e}ron
(1995, 1996) and Hawkins (1996).  The selection of subsamples of quasars in
different redshift bands typically used colour limits together with variability
and compactness criteria, but in all cases the quasar images were required to
be round.  This was to eliminate overlapping images which might masquerade as
quasars, or even quasars merged with stars or galaxies where the photometry
would give misleading results.  One consequence of this was to reject any
gravitationally lensed quasars from consideration, where the split image would
appear elongated.  To rectify this a search was designed specifically to
look for gravitationally lensed systems.  The requirements were that the images
should have a major to minor axis ratio greater than 1.5, and an ultra-violet
excess $U-B < -0.4$.  Of the 200,000 objects in the field, 500 have $U-B < -0.4$
of which 23 were elongated.  Of these, 12 were variable according to
our usual criteria (Hawkins \& V\'{e}ron, 1995), with an amplitude greater than
0.35 mag.  4 of the sample were quasars with overlapping galaxies previously
found by Morris {\em et al.} (1991), and so we set about obtaining spectra for
the remainder, several of which appeared to be excellent candidates for lensed
systems.  It also seems possible that some lensed quasars will be found among
the non-variable objects.

\vspace{1.0cm}

\noindent
{\bf 2.2 The Double Quasar Q2138-431}

\vspace{0.3cm}

  The first candidate to be studied in detail appeared as two star-like
images separated by 4.5 arcseconds.  Spectra of the two components were
obtained on the ESO 3.6m telescope at La Silla by aligning the slit along the
line of centres.  The spectra were very similar, and the redshifts appeared to
be the same, $z = 1.64$.  There seemed to be a {\em prima facie} case for a
gravitationally lensed system and so a few nights later we obtained a second
higher signal-to-noise observation in both blue and red wavelength bands
covering a combined spectral range from 3700\AA \ to 10000\AA.  The spectra are
shown in Fig. 1, and it will be seen that they closely resemble each other.
Fig 2. shows the quotient of the two spectra which is almost flat, implying
no significant differences.\\

  The redshifts were first calculated by measuring the emission lines in each
spectrum, which gave $z = 1.638 \pm 0.004$ and $z = 1.644 \pm 0.005$ for
the two components, the same within the errors.  In order to
obtain a more accurate measurement a cross-correlation routine was applied.
This gave identical redshifts within the errors, and a velocity difference
of $0 \pm115$ km.sec$^{-1}$.\\

  Optical photometry of the system was obtained with a CCD camera on the
ESO 2.2m telescope at La Silla in August 1995.  The pixel scale was
0.336 arcsec/pixel and the seeing averaged 1.2
arcsecs which allowed a clear separation of the images.  The R-band frame
is shown in Fig. 3, where the faintest visible objects have magnitude
$R > 24$.  The results are shown in Table 1, where the 5 columns are the
colour passband, the apparent magnitude of the
\begin{table}[h]
\centering
\caption{CCD Photometry for Q2138-431AB}
\vspace{10mm}
\begin{tabular}{cccll}
colour&$m_A$&$m_B$&$f_A/f_B$&$m_{gal} >$\\
&&&&\\
 $B$ & 21.02 & 19.83 & 0.33  & 23. \\
 $V$ & 20.85 & 19.57 & 0.307 & 23.5\\
 $R$ & 20.43 & 19.18 & 0.318 & 23.8\\
 $I$ & 20.12 & 18.86 & 0.313 & 22.8\\
\end{tabular}
\end{table}
two components, the flux ratio and the lower limit to the magnitude
of any lensing galaxy.  It will be seen that at the time of
observation the magnitude difference between the two components
was $\delta m = 1.2$ with no significant dependence on colour.  Fig. 4
shows the $V-R$ and $R-I$ colours for the two components of Q2138-431
together with photographic measures for 10 other quasars from the sample
with very similar redshift.  The colours of the double quasar are identical
within the errors of the CCD observations, whereas the other 10 quasars
in the sample illustrate the wide range of colours which quasars at this
redshift can exhibit.

\vspace{1.0cm}

\noindent
{\bf 2.3 Search for the Lensing Galaxy}

\vspace{0.3cm}

  To fully understand a gravitational lens system, and more specifically to
use it to measure the Hubble constant, the lensing galaxy must be located.
It is then necessary to measure its redshift and estimate the mass distribution
relative to the quasar images.  In fact, of the seven double quasar systems
so far discovered with image separation greater than 2 arcseconds, lensing
galaxies have only been found for two.\\

  To detect the lensing galaxy for Q2138-431 we used the deep CCD frames
described above for the photometric measurements.  Initial examination
of the area around the quasar system showed no objects which might act
as gravitational lenses.  We obtained a more useful limit by using stars in
the vicinity of the quasar images to obtain an accurate measure of the
point spread function (PSF).  This was then subtracted from each of the
quasar images in the hope of revealing an underlying lensing galaxy.
There is an element of uncertainty in the normalisation of the PSF 
in this procedure, but in the event we found that both quasar
images subtracted out exactly.  To put an upper limit on the magnitude
of a possible lensing galaxy, we extracted the faint galaxy visible to
the southeast of the quasar in Fig. 3.  We then placed it at various
points between the two quasar images, varied its brightness and carried
out the PSF subtraction procedure.  This enabled us to put an upper limit of
$R > 23.8$ for a potential lensing galaxy.  We also obtained a K-band
image of the field with the IRAC2 infrared camera on the ESO 2.2m.
A 5 hour integration failed to reveal a lensing galaxy between the two images,
although there was evidence for additional K-band flux associated with the
brighter component.  This observation raises the possibility that a very
red lensing galaxy may be lying close to the brighter quasar image, a
configuration which requires fine tuning of the model parameters.\\

  One can now ask what limits can be put on the mass-to-light ratio
of a lensing galaxy capable of producing the observed image splitting and
flux ratio, but constrained to be fainter than the observed magnitude limit.
We have modelled the system assuming both a point mass and a more realistic
galaxy profile, and have thus derived a lower limit to the mass-to-light
ratio as a function of redshift.   If the lens can be modelled as a point
mass, the brightness ratio $R>1$ of the two images and the separation 
$\Delta\theta$ on the 
sky can be used to calculate the Einstein radius $\theta_E = 
\sqrt{4GMD_{LS}/(D_{OL}D_{OS}c^2)}$, where $M$ is the lens mass, and the $D$
are the angular diameter distances between observer, lens and source.  It
is straightforward to show, from the lensing equations (see e.g. Schneider 
et al. 1992) that $\theta_E=\sqrt{1-f^2}\Delta\theta/2$, where 
$f\equiv(R+1-\sqrt{4R})/(R-1)$.   $\theta_E$ is 2.1 arcsec, and the required
mass is shown in Fig. 5(a), for two different cosmologies 
(solid line: Einstein-de Sitter, dashed line: $\Omega_0=0.1$).   
Also shown (dot-dashed) is the mass required in a more realistic 
Hernquist mass profile (Hernquist 1990),
with a density run $\rho(r) = M/\left[2\pi r_c^3 s(1+s)^3\right]$ 
where $s=r/r_c$ and the
core radius is taken to be $r_c= 1.7 h^{-1}$ kpc.  An advantage of this 
profile is that the bending angle may be written in closed form;  the 
enclosed mass within a projected radius $rr_c$ is 
\begin{eqnarray}
{M(<rr_c)\over M} & = & 
{r^2\over r^2-1} - {r^2\over (r^2-1)^{3/2}} {\rm cos}^{-1}
\left({1\over r}\right)\qquad r>1\\
& = & {r^2\over r^2-1} - {r^2\over (1-r^2)^{3/2}}{\rm ln}\left({r\over
1-\sqrt{1-r^2}}\right)\qquad r<1
\end{eqnarray}
with $M(<r_c)=M/3$.  Here one needs to search
for a solution with the correct brightness ratio and separation, and
we see from Fig. 5(a) that for $r_c$ appropriate for galaxies,
the required mass is similar to the point mass calculation.      
In view of the relatively large
masses required if the lens is near the source, it is worth exploring a larger 
core radius, appropriate for a cluster.   However, the only Hernquist 
profiles which are able to
produce split images with the required amplification ratio have core radii
less than 7 $h^{-1}$ kpc, so we are restricted to galaxy-like objects
(Fig. 5(a) also shows the mass required for a core radius of 5 $h^{-1}$ kpc).
It is also worth noting that it requires an astonishing 
degree of fine tuning for the faint third image in the Hernquist model
to alter significantly the brightness ratio by merging with another image.\\

It will be seen from Fig. 5(a)
that there is an absolute lower limit of $200M_{\odot}/L_{\odot}$ when
the lens is at a redshift $z = 1.5$, for an Einstein-de Sitter universe, and 
a slightly lower limit if the universe is open.  In fact this configuration 
is highly improbable, and the most likely position for the lens is at around
$z = 0.5$ (Turner {\em et al.}, 1984), implying a minimum mass-to-light ratio
of $1000M_{\odot}/L_{\odot}$.  In this case the lensing object would
presumably be some form of `dark galaxy', or perhaps a dark matter galactic
halo.\\

Another approach is to consider the possible effect of shear or convergence
produced by a nearby group or cluster of galaxies.  This is a model which
has been used to describe Q0957+561 (Bernstein {\it et al.}, 1993), and
also for the wide separation lens Q2345+007 (Pell\'{o} {\it et al.}, 1996).
In both of these cases a group of galaxies is detected close to the quasar
images, and the splitting is attributed to the resulting shear field.
We are in a much worse position here to model such a situation, as we cannot
even establish where the lensing galaxy is (if present).  The
addition of a uniform screen, representing a smooth cluster, would reduce the
mass estimates by a factor $1-\Sigma/\Sigma_c$, where the critical surface
density is $\Sigma_c = c^2 D_{OS}/(4\pi G D_{OL}D_{LS}$.  The mass requirement
is thus eased if the cluster has a substantial fraction of the critical
surface density.   Numerical modelling of clusters (Bartelmann \& Weiss 1994) 
indicates that this may be possible, but modelling of observed arcs
routinely requires a contribution from a central galaxy (Miralda-Escude 
\& Babul 1995).   Although a significant cluster contribution remains an 
open possibility,  the very small core radius required argues 
against it, and there is no evidence apparent in the images:  
fig. 6 shows an area of approximately 8 arcminutes on
a side centered on the double quasar.  The frame is taken from a digital stack
(Hawkins, 1994) of 64 UK Schmidt plates in the IIIa-J/GG395 passband with
effective wavelength 4500 \AA.  The limiting magnitude is $B_J \sim 24$, and
there is no sign of a cluster within 2 arcminutes of the quasar.  In fact
judging by the surrounding background the system lies in a particularly clear
region of sky, the nearest cluster being in the top right hand corner of the
field.

\vspace{2cm}

\noindent
{\Large\bf 3. DISCUSSION}
\vspace{0.5cm}

  The properties of the double quasar Q2138-431AB may be summarised as
follows:\\

  Redshift $z = 1.641$, velocity difference $\delta v = 0\pm115$ km.sec$^{-1}$\\
\indent
  Separation $= 4.5''$\\
\indent
  $B$ magnitude $m_A = 19.8$, $m_B = 21.0$\\
\indent
  Variability amplitude $\delta m_A = 1.1$, $\delta m_B = 0.6$\\

\noindent We now address the question of the underlying nature of the system.
There seem to be three possibilities:\\

\noindent
1.  A chance association of two separate quasars, possibly made more likely
by the effects of clustering.\\

\noindent
2.  A pair of quasars in a bound orbit, forming a binary system.\\

\noindent
3.  A single quasar gravitationally lensed by a dark galaxy or galactic halo
with a mass around $10^{12} - 10^{13} M_{\odot}$.\\

  The likelihood of a chance coincidence can first be assessed by considering
the surface density of the parent population of quasars in the field and
asking what is the probability $P$ that two will lie within 4.5 arcsecs of each
other.  The parent population of single quasars with similar characteristics
to the lens candidates comprised 310 objects with $U-B < -0.4$ and $B < 21$ in
an area of 18.8 square degrees.  This gives a surface density of quasars of
about 16 per square degree, which implies
a probability of about 1 percent for any companions within $4.5''$, 
for the parent sample of $\sim$310 quasars.  In practice one would expect 
this figure to be modified by clustering, which enhances the probability,
and by the redshift information, which reduces it.   The small-scale 
clustering of quasars is poorly constrained, but assuming that it follows
the galaxy correlation function $\xi(r) \simeq (r/r_0)^{-\gamma}$,
with $r_0 \simeq 5 h^{-1}$ Mpc and $\gamma \simeq 1.8$ (Collins {\it et al.},
1992, Vogeley {\it et al.}, 1992), boosted by
a relative bias $b_Q^2$, then we can calculate the probablity of a pair 
within $50 h^{-1}$ kpc (the comoving separation corresponding to $4.5''$ 
at z=1.6 if $\Omega_0=1$).  
Using the comoving number density of $1.7\times 10^{-5} h^3$ Mpc$^{-3}$ 
obtained from
the redshift distribution in the sample at redshifts around 1 to 1.5,
we find this probability is about 1.5\% if $b_Q=1$ and the clustering does
not evolve.  This is, however, an
underestimate for lensing candidates, since the radial separation can
far exceed $50 h^{-1}$ kpc and still be considered a good lensing
candidate.  For illustration, a velocity difference of 100 km s$^{-1}$ 
at a redshift of 1.6 corresponds to a comoving separation of around 
600 $h^{-1}$ kpc in an Einstein-de-Sitter Universe, and this could easily
be lost in the uncertainties of redshift determination.  If instead of a
radial separation of 50 $h^{-1}$ kpc we use this larger figure of 600 $h^{-1}$
kpc, it would increase the probability to around 2.5\%.  Peculiar velocities
modify this in a model-dependent manner, reducing the chance of good 
agreement in the redshifts for virialised systems and increasing it for
collapsing systems, but the point is that the probability of a close
separation in angular and redshift terms is small but not negligible.\\

  The idea that the quasars form part of a gravitationally bound binary system
is much harder to test, mainly because the circumstances surrounding the
formation and evolution of such a system are largely a matter of
speculation.  There are a number of observations which appear to count against
this possibility, such as the extreme similarity of the spectra and colours of
the two components and the small differential velocity.  It is also clear from
the discussion in the previous paragraph that from a statistical point of
view, binary quasars might just be part of a clustering hierarchy.  
However, none of these arguments are sufficient to rule
out the essentially unconstrained concept of binary quasars, and it
must remain a possible explanation.\\

  A plausible justification for the binary, or clustered, model, is
that quasar activity may well be triggered by a close encounter between
two galaxies,  with a small probability of triggering.   The possibility
then arises that, in rare cases, quasar activity may be initiated in both
galaxies in the encounter.  The explanation of the similarity of the 
spectra, which are not absolutely identical,  might then lie in the fact 
that the quasars would have the same formation epoch, and would be observed 
at the same time after formation.  As part of a common system, their 
abundances might not significantly differ.\\

  The third possibility, that the system is gravitationally lensed, is well
supported by most of the available observations.  The similarity of the spectra
and colours, and the small velocity difference between the two components
are all to be expected from a gravitationally lensed system.  The problem is
the failure to find the lensing galaxy.  Given the apparent absence of a shear
field, this means that to make a case for a gravitationally lensed system one
must postulate a dark galaxy as the lens.  Although this may seem a radical
step, it is in fact a position which several other groups have adopted when
analysing double quasar systems (eg Tyson {\it et al.}, 1986).
An obvious possibility is that the lensing object is a low surface brightness
galaxy,  which fails not because of the surface brightness limit corresponding
with the R limit of 23.8, but rather because 
known low surface brightness galaxies do not have the very large mass-to-light
ratio required (Sprayberry et al. 1995).  We are left with the conclusion that,
unpalatable though the idea of dark galaxies may be, it seems to promise the
most plausible explanation for the observations.

\vspace{2cm}

\noindent
{\Large\bf 4 CONCLUSIONS}
\vspace{0.5cm}

  We have reported the discovery of a new double quasar Q2138-431 which we have
observed in some detail with a view to establishing whether it is a
gravitationally lensed system.  It comprises two images with magnitudes
$B = 19.8$ and $B = 21.0$ separated by 4.5$''$.  The spectra and colours of the
two components are very similar, with a redshift of $z = 1.461$ and velocity
difference $\delta v = 0 \pm 115$ km.sec$^{-1}$.  In spite of an intensive
search we have failed to find a lensing galaxy, which must have a magnitude
$R > 23.8$.  This has enabled us to put a lower limit on the mass-to-light
ratio of any lensing galaxy in the range 200 to 1000 $M_{\odot}/L_{\odot}$
depending on its redshift.  We have considered three possible interpretations
of the system:\\

\noindent
1.  Chance coincidence of the two images, which we rule out on statistical
grounds.\\

\noindent
2.  A gravitationally bound binary quasar.  The similarity of the two
components and the small velocity difference count against this possibility,
but we feel we cannot rule it out.\\

\noindent
3.  A gravitational lens.  Most of the observations favour this picture, but
the lensing object would have to be a dark galaxy or dark matter halo.  This
would clearly require a departure from the conventional idea of the galaxy
population.\\

\vspace{2cm}

\noindent
{\Large\bf REFERENCES}
\vspace{1.5cm}

\ref Bartelmann M., Weiss A., 1994, A \& A 287, 1
\ref Bernstein G.M., Tyson J.A., Kochanek C.S., 1993, AJ, 105, 816
\ref Collins C.A., Nichol R.C., Lumsden, S.L., 1992, MNRAS, 254, 295
\ref Hawkins M.R.S., 1993, Nat, 366, 242
\ref Hawkins M.R.S., 1994, in {\it Astronomy from Wide-Field Imaging} (ed
MacGillivray), IAU Symp. 161, p. 177
\ref Hawkins M.R.S., 1996, MNRAS, 278, 787
\ref Hawkins M.R.S., V\'{e}ron P., 1995, MNRAS, 275, 1102
\ref Hawkins M.R.S., V\'{e}ron P., 1996, MNRAS, 281, 348
\ref Hernquist L., 1990, ApJ, 356, 359
\ref Miralda-Escude J.,  Babul A., 1995, ApJ 449, 18
\ref Morris S.L., Weymann R.J., Anderson S.F., Hewett P.C., Foltz C.B.,
Chaffee F.H., Francis P.J., MacAlpine G.M., 1991, AJ, 102, 1627
\ref Pell\'{o} R., Miralles J.M., Le Borgue J.-F., Picat J.-P., Soucail G.,
Bruzual G., 1996, A\&A, 314, 73
\ref Refsdal S., 1964, MNRAS, 128, 307
\ref Schneider P., Falco E.E., Ehlers J.L., 1992, `Gravitational Lenses', 
Springer, Berlin
\ref Sprayberry D., Bernstein G.M., Impey C.D., 1995, ApJ, 438, 72
\ref Turner E.L., Ostriker J.P., Gott J.R., 1984, ApJ, 284, 1
\ref Tyson J.A., Seitzer P., Weymann R.J., Foltz C., 1986, AJ, 91, 1274
\ref Vogeley M.S., Park C., Geller M.J., Huchra J.P., 1992, ApJ, 391, L5
\ref Walsh D., Carswell R., Weymann R., 1979, Nat, 279, 381

\newpage

\vspace{2cm}

\noindent
{\Large\bf FIGURE CAPTIONS}
\vspace{1.5cm}

\noindent
{\large\bf Figure 1.} \\

\noindent
Spectra for the two images of the double quasar Q2138-431AB in red and blue
passbands.

\vspace{0.5cm}

\noindent
{\large\bf Figure 2.} \\

\noindent
Spectrum of component A divided by that of component B from Fig. 1.

\vspace{0.5cm}

\noindent
{\large\bf Figure 3.} \\

\noindent
Part of an R-band CCD frame of the field around the double quasar
Q2138-431.  The plot is 50 arcseconds on a side, and North is up
the page, East to the left.  The centroid of the two quasar images
is at 21h 38m 6.66s, -43$^{\circ}$ 10$'$ 50.0$''$ (1950), and they
are separated by 4.5 arcseconds.  The star about 20 arcseconds to
the South was used for the image subtraction, the effect of which
is illustrated in the second panel.

\vspace{0.5cm}

\noindent
{\large\bf Figure 4.} \\

\noindent
The $V-R$ vs $R-I$ relation for quasars with $z \approx 1.64$.  The two
components of Q2138-431 are shown as open circles, and other quasars from
the field 287 sample as closed circles. 

\vspace{0.5cm}

\noindent
{\large\bf Figure 5.} \\

\noindent
(a) The mass of the postulated lens for the double quasar
Q2138-431AB as a function of the lens redshift.  The solid
line assumes a point mass lens and an Einstein-de Sitter universe;  the dashed
line assumes that $\Omega_0=0.1$ and no vacuum energy.  The dot-dashed line
shows the required mass if the density distribution follows a Hernquist 
profile with core radius 1.7 $h^{-1}$ kpc (corresponding with an
effective radius of 3 $h^{-1}$ kpc) and the dotted line assumes a core
radius of 5 $h^{-1}$ kpc.  The last two curves assume $\Omega_0=1$.  
(b) The minimum mass-to-light ratio for a lens for the double quasar
Q2138-431AB.  Lines as in (a).

\vspace{0.5cm}

\noindent
{\large\bf Figure 6.} \\

\noindent
Plot of the area centered on Q2138-431.  The frame is approximately 8
arcminutes square and is derived from digitally stacked photographic
plates with a limiting magnitude of $B_J \sim 24$.  North is up the
page, East to the left.

\begin{figure}
     \setlength{\unitlength}{1mm}
     \begin{picture}(50,180)
     \put(0,0){\includegraphics{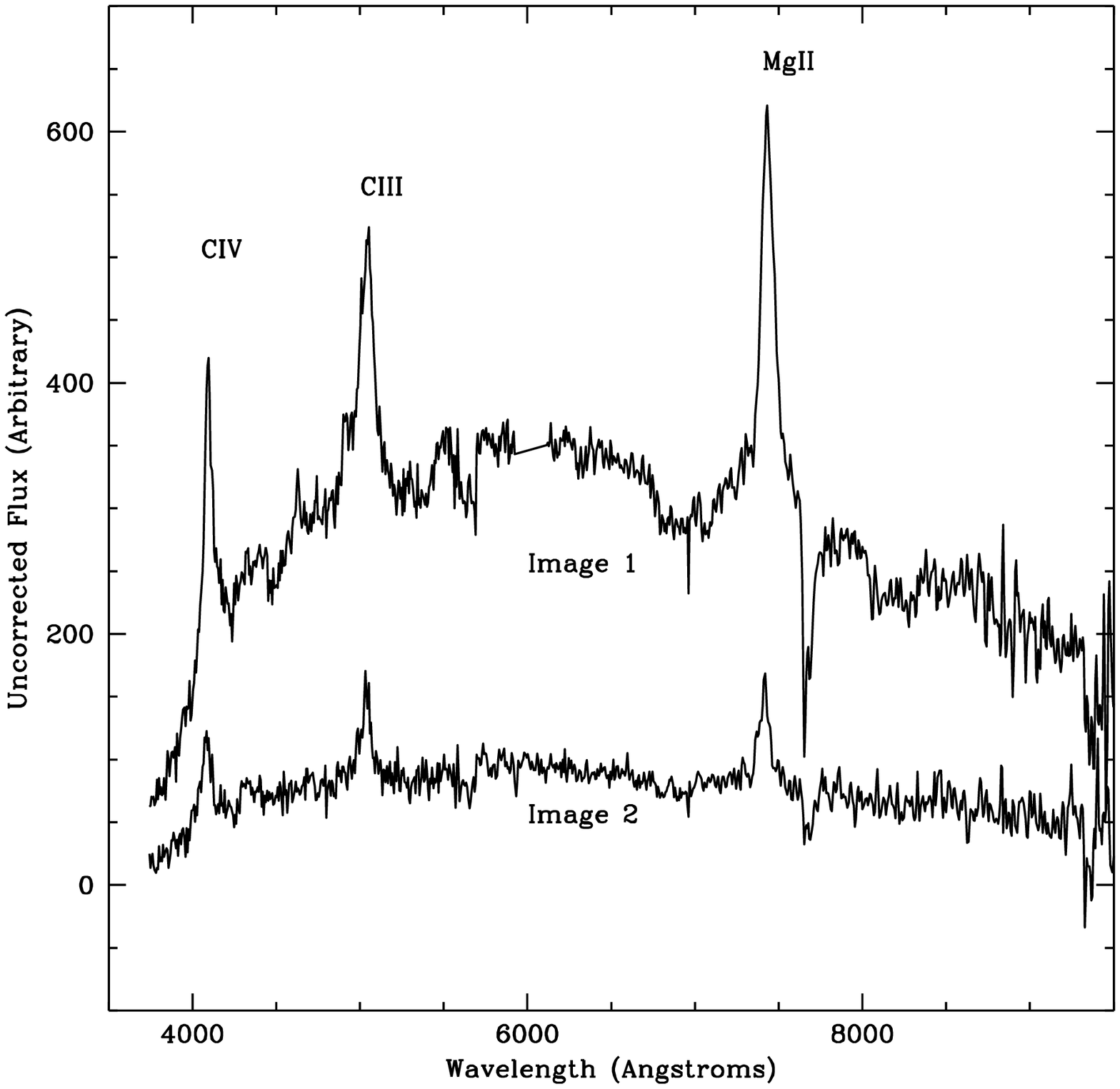}}
     \end{picture}
%     \caption{}
\end{figure}
\suppressfloats
\begin{figure}
     \setlength{\unitlength}{1mm}
     \begin{picture}(50,180)
     \put(0,0){\includegraphics{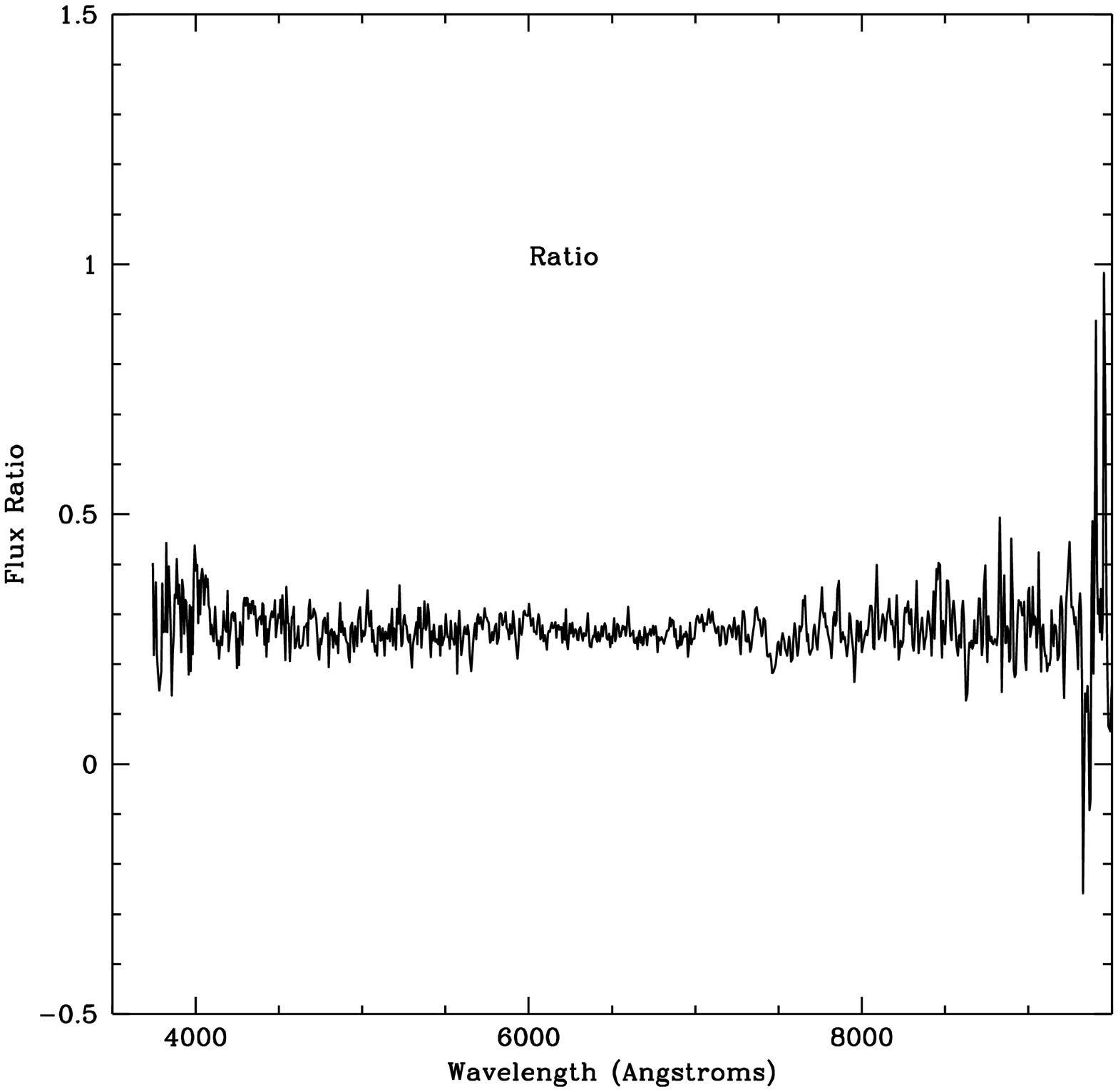}}
     \end{picture}
%     \caption{}
\end{figure}
\suppressfloats
\newpage
%\phantom{.}
%\newpage
\begin{figure}
     \setlength{\unitlength}{1mm}
     \begin{picture}(50,150)
     \put(0,0){\includegraphics{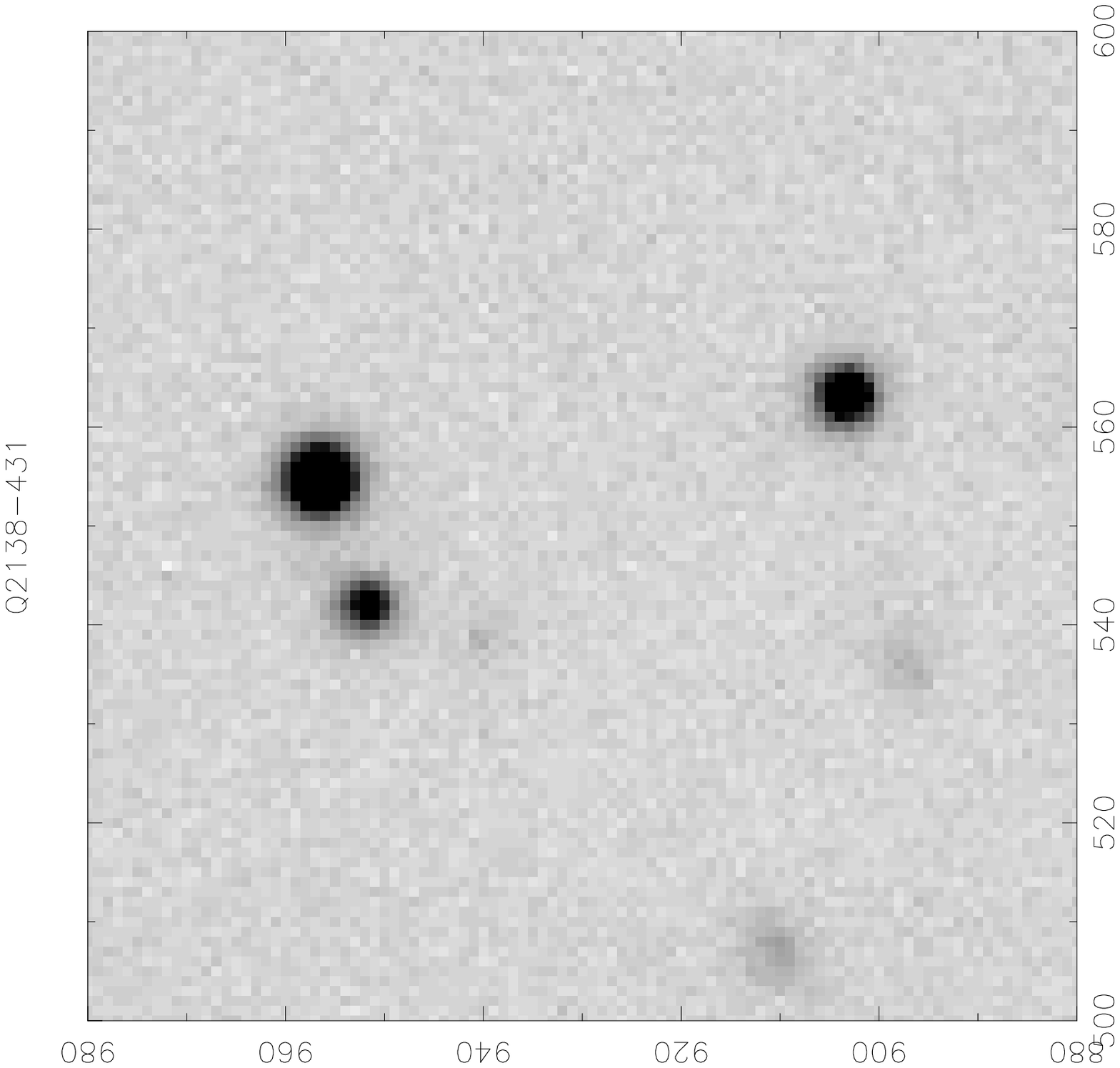}}
     \end{picture}
%     \caption{}
\end{figure}
\suppressfloats
\newpage
\begin{figure}
     \setlength{\unitlength}{1mm}
     \begin{picture}(50,150)
     \put(0,0){\includegraphics{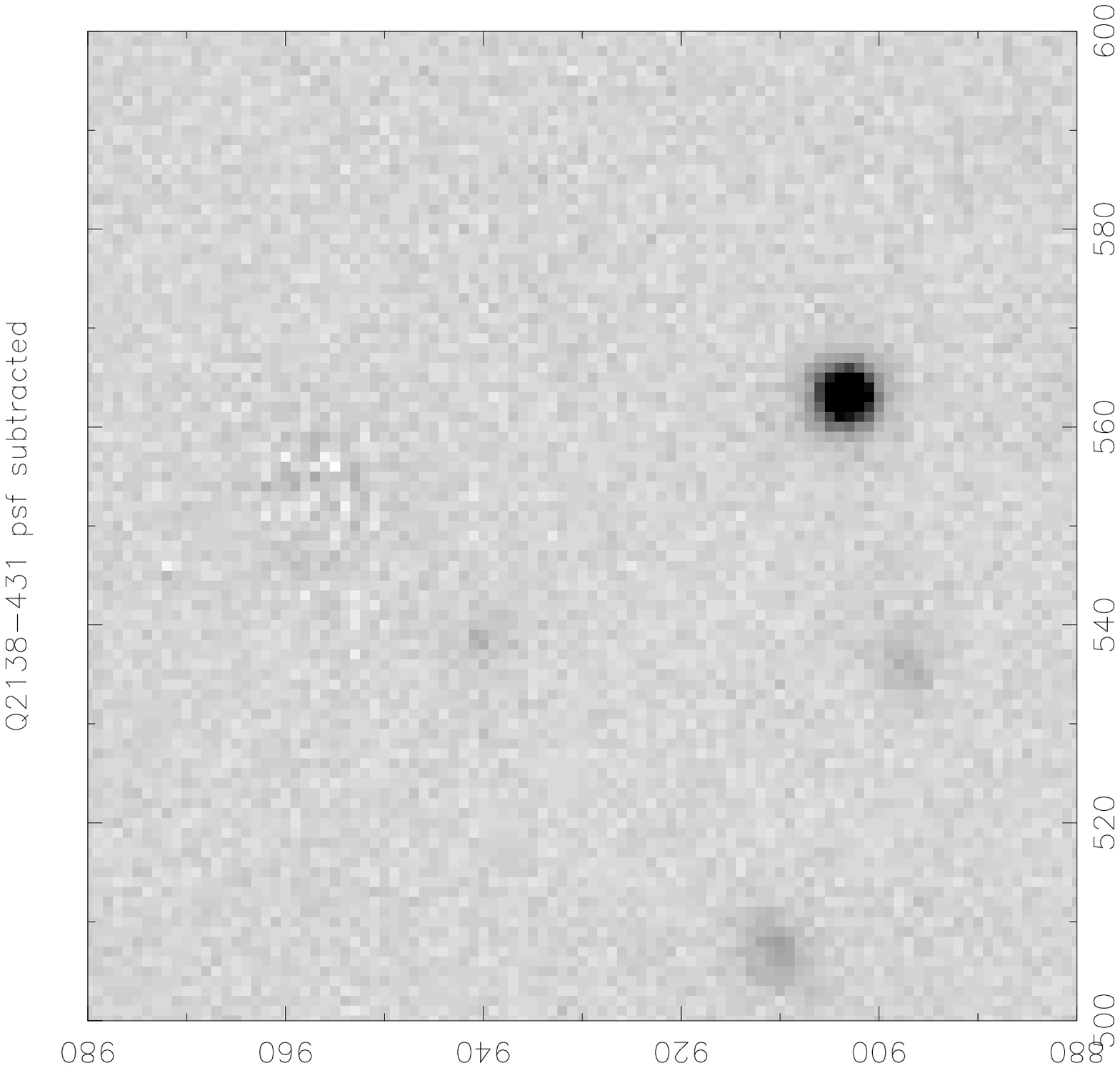}}
     \end{picture}
%     \caption{}
\end{figure}
\suppressfloats
\newpage
\begin{figure}
     \setlength{\unitlength}{1mm}
     \begin{picture}(50,150)
     \put(0,0){\includegraphics{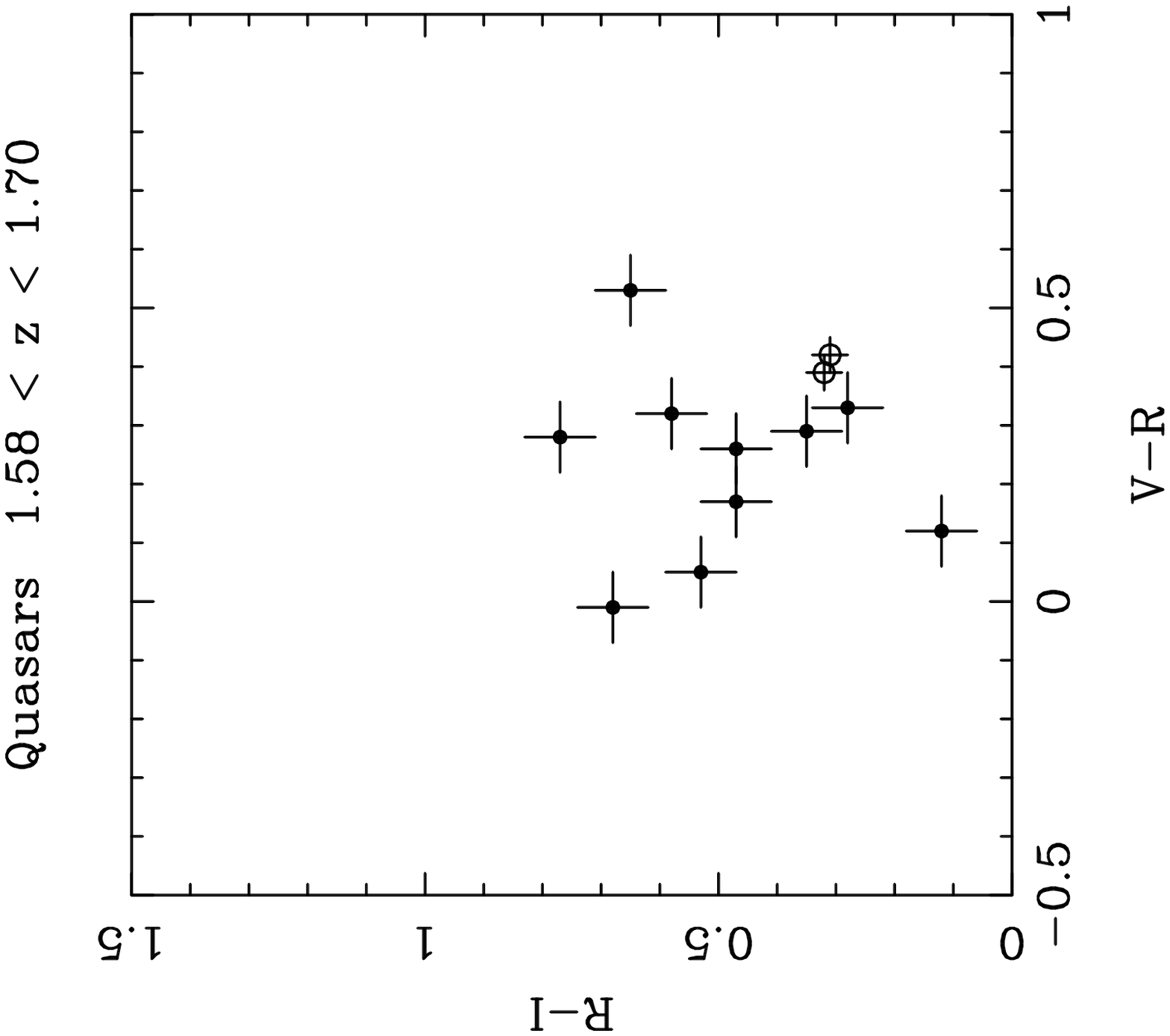}}
     \end{picture}
%     \caption{}
\end{figure}
\suppressfloats
\newpage
\begin{figure}
     \setlength{\unitlength}{1mm}
     \begin{picture}(50,150)
     \put(0,0){\includegraphics{fig5a.ps}}
     \end{picture}
%     \caption{}
\end{figure}
\suppressfloats
\newpage
\begin{figure}
     \setlength{\unitlength}{1mm}
     \begin{picture}(50,150)
     \put(0,0){\includegraphics{fig5b.ps}}
     \end{picture}
%     \caption{}
\end{figure}
\suppressfloats
\newpage
\begin{figure}
     \setlength{\unitlength}{1mm}
     \begin{picture}(50,150)
     \put(0,0){\includegraphics{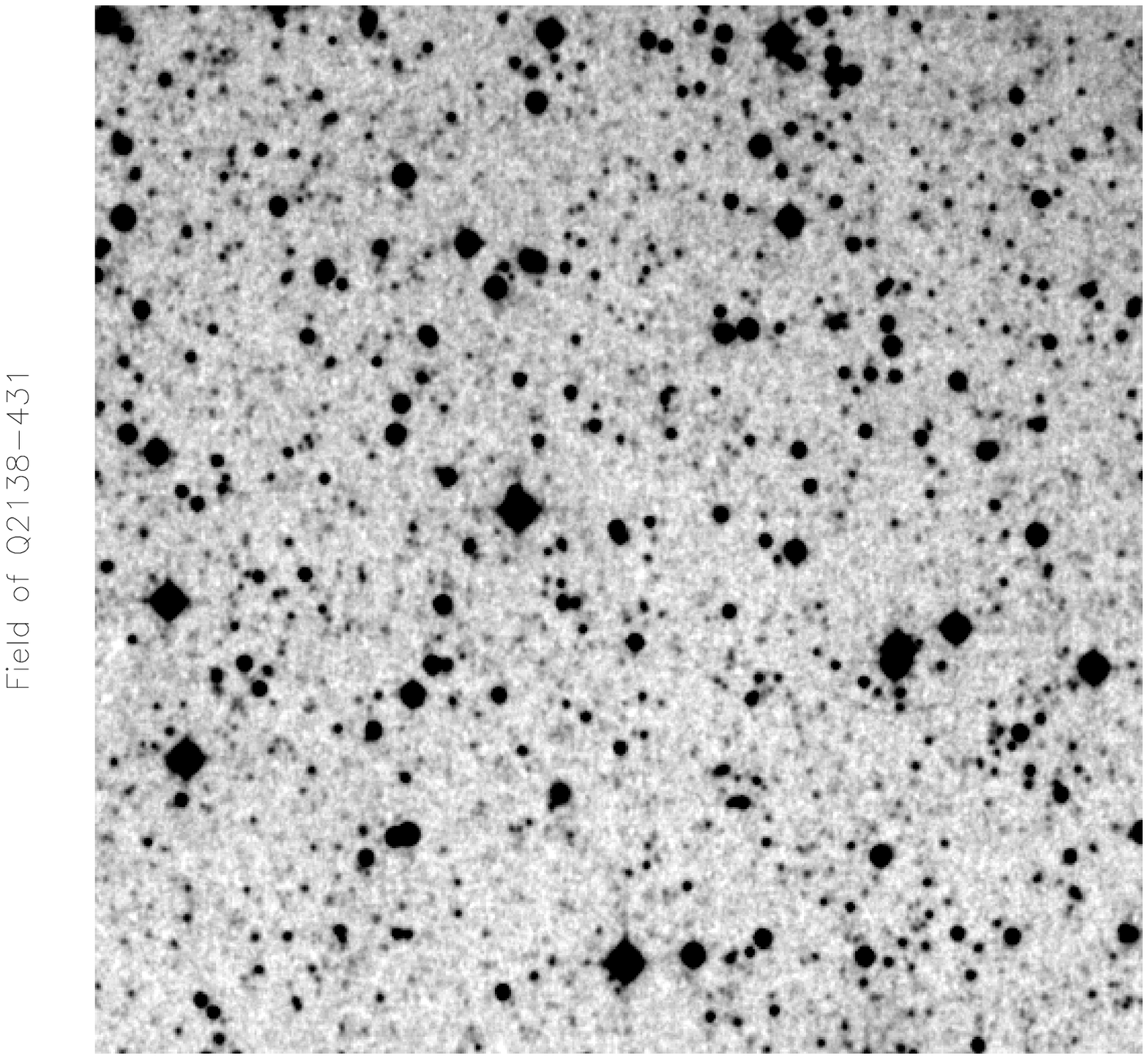}}
     \end{picture}
%     \caption{}
\end{figure}

\end{document}